\documentclass[conference]{IEEEtran}
\usepackage{cite}
\usepackage{amsmath,amssymb,amsfonts}
\usepackage{algorithmic}
\usepackage{graphicx}
\usepackage{textcomp}
\usepackage{xcolor}
\usepackage{float}

\usepackage{tikz}

\newcommand\acceptedtext{%
  \footnotesize Accepted for IEEE 7th World Forum on Internet of Things 2021}
\newcommand\acceptednotice{%
\begin{tikzpicture}[remember picture,overlay]
\node[anchor=north,yshift=-20pt] at (current page.north) {\fbox{\parbox{\dimexpr\textwidth-\fboxsep-\fboxrule\relax}{\acceptedtext}}};
\end{tikzpicture}%
}

\newcommand\copyrighttext{%
  \footnotesize © 2021 IEEE. Personal use of this material is permitted. Permission from IEEE must be obtained for all other uses, in any current or future media, including reprinting/republishing this material for advertising or promotional purposes, creating new collective works, for resale or redistribution to servers or lists, or reuse of any copyrighted component of this work in other works. DOI: 10.1109/WF-IoT51360.2021.9595119}
\newcommand\copyrightnotice{%

\begin{tikzpicture}[remember picture,overlay]
\node[anchor=south,yshift=20pt] at (current page.south) {\fbox{\parbox{\dimexpr\textwidth-\fboxsep-\fboxrule\relax}{\copyrighttext}}};
\end{tikzpicture}%
}

\def\BibTeX{{\rm B\kern-.05em{\sc i\kern-.025em b}\kern-.08em
    T\kern-.1667em\lower.7ex\hbox{E}\kern-.125emX}}

\begin{document}

\title{IoT Virtualization with ML-based Information Extraction}

\author{\IEEEauthorblockN{Martin Bauer}
\IEEEauthorblockA{\textit{NEC Laboratories Europe} \\
Heidelberg, Germany \\
martin.bauer@neclab.eu}
}

\maketitle
\acceptednotice

\begin{abstract}
For IoT to reach its full potential, the sharing and reuse of information in different applications and across verticals is of paramount importance. However, there are a plethora of IoT platforms using different representations, protocols and interaction patterns. To address this issue, the Fed4IoT project has developed an IoT virtualization platform that, on the one hand, integrates information from many different source platforms and, on the other hand, makes the information required by the respective users available in the target platform of choice. To enable this, information is translated into a common, neutral exchange format. The format of choice is NGSI-LD, which is being standardized by the ETSI Industry Specification Group on Context Information Management (ETSI ISG CIM). Thing Visors are the components that translate the source information to NGSI-LD, which is then delivered to the target platform and translated into the target format.

ThingVisors can be implemented by hand, but this requires significant human effort, especially considering the heterogeneity of low level information produced by a multitude of sensors. Thus, supporting the human developer and, ideally, fully automating the process of extracting and enriching data and translating it to NGSI-LD is a crucial step. Machine learning is a promising approach for this, but it typically requires large amounts of hand-labelled data for training, an effort that makes it unrealistic in many IoT scenarios. A programmatic labelling approach called knowledge infusion that encodes expert knowledge is used for matching a schema or ontology extracted from the data with a target schema or ontology, providing the basis for annotating the data and facilitating the translation to NGSI-LD.
\end{abstract}


\copyrightnotice

\section{Introduction}
With the proliferation of mobile and embedded computing devices with sensing and communication capabilities, the Internet of Things (IoT) has entered many areas of our daily life. However, in reality, there is not yet one connected \emph{Internet} of Things, but rather heterogeneous, disconnected areas based on different technologies and applications that are tightly coupled to specific deployments. For IoT to reach its full potential, the sharing and reuse of information~\cite{iiot2018datasharing} in different applications and across verticals is of paramount importance.
The different IoT technologies and platforms will not disappear and magically merge into a single one in the near term future. Also, they each have their pros and cons and developers and information providers have their preferred IoT platforms. To nevertheless support the sharing and reuse of information and thus increase the value of IoT deployments and applications, the interoperability between different IoT systems has to be achieved.

Device virtualization has been successful in providing a virtual computing environment, independent of the actual underlying physical hardware, which can be flexibly deployed. Today this forms the basis of cloud computing, enabling users to dynamically deploy their applications and scale as needed. The idea is to apply the same approach to the Internet of Things, i.e. making IoT information and resources available in virtualized form, according to the needs of the users.

\begin{figure}[!htbp]
  \centering
  \includegraphics[width=3.4in]{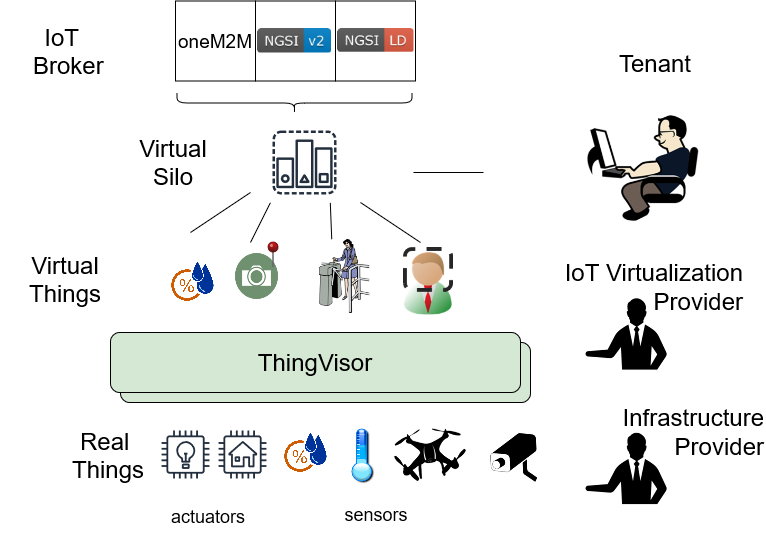}
  \caption{IoT virtualization}
  \label{fig:iot_virtualization}
\end{figure}

\begin{figure*}[!htbp]
  \centering
  \includegraphics[width=7in]{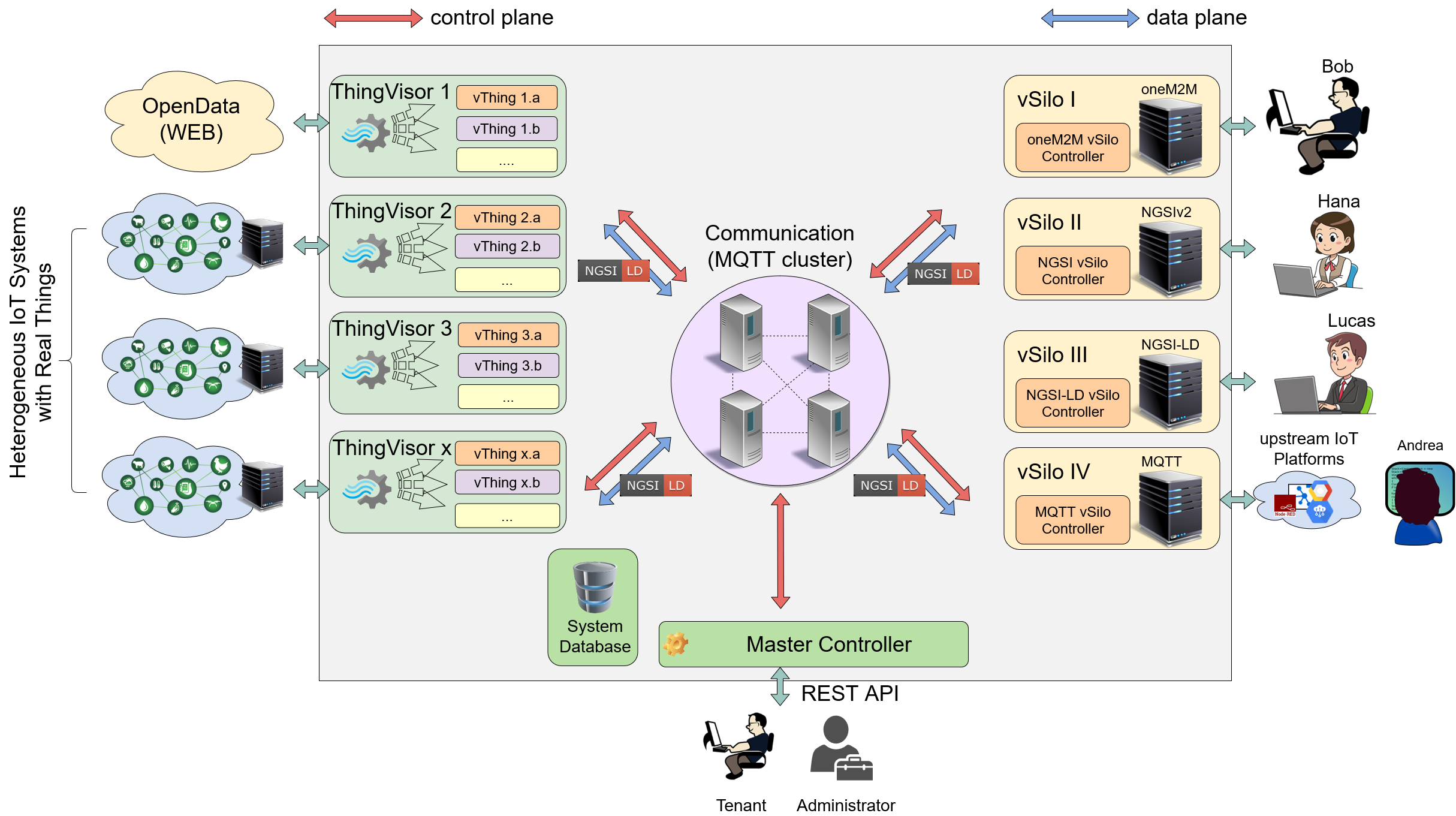}
  \caption{VirIoT system architecture}
  \label{fig:viriot_system_architecture}
\end{figure*}

In the EU-Japan research project Fed4IoT~\cite{fed4iot}, we have developed the VirIoT virtualization platform~\cite{detti2019IoTSystemVirtualization}, which is described in Section~\ref{sec:iot_virtualization_platform}. It enables making IoT information from many source platforms available in the desired target platform, as needed by the respective user. To enable this, a common neutral information format is required, for which the NGSI-LD information model was chosen. NGSI-LD is specified by the ETSI Industry Specification Group on Context Information Management (ETSI ISG CIM) and is introduced in Section~\ref{sec:ngsi-ld}. While the VirIoT platform enables the sharing of information, significant human effort is still needed for coding the translation from a variety of source formats to the neutral exchange format. Automating this process, or at least supporting the developers in this undertaking, is essential to make this practically viable in many real-world scenarios. In Section~\ref{sec:ml-based_information_extraction}, we explore the different steps for enabling this translation. In particular, we focus on one key step, how the conceptual source model can be matched with the target model, assuming that both are represented as ontologies. A weakly supervised machine-learning approach called \emph{Knowledge Infusion} is proposed for the matching of ontologies. Finally, Section~\ref{sec:conclusions_and_outlook} provides a conclusion and an outlook.

\section{IoT Virtualization Platform}
\label{sec:iot_virtualization_platform}

Virtualization has enabled the success of cloud computing as it decouples infrastructure providers from application providers. In IoT, there is currently a strong coupling between IoT infrastructures and applications, which results in IoT silos. The idea is to bring the same kind of decoupling to the IoT world that has been successful in cloud computing.

In Figure~\ref{fig:iot_virtualization}, the basic idea behind the IoT virtualization is shown. Infrastructure providers make real things, i.e. sensors and actuators available. Based on the concept of a ThingVisor, an IoT virtualization provider exposes virtual things - similar to cloud providers using a hypervisor to provide virtual hardware. In the simplest case, virtual things provide the same functionality as the real things, e.g. a real temperature sensor can be made available as a virtual temperature sensor, but there can also be additional processing, e.g. instead of directly providing a virtual camera based on a real camera, virtual thing visor can provide the count of faces or the people recognized on the current frame from the camera. Due to limited available space and the focus of this paper, we cover only  sensor-based vThings in the following. The combination of virtual things needed by the user, in this case called a \emph{tenant}, is exposed as a virtual silo - similar to exposing a virtual machine in the cloud. In each virtual silo an IoT broker chosen by the tenant, e.g. an implementation of oneM2M~\cite{datta2015oneM2M}, FIWARE NGSIv2 or NGSI-LD~\cite{ngsild2021apispecification} can be made available, enabling the tenant to access the virtual thing using the API and underlying information model of choice. This is similar to choosing the operating system for a virtual machine running in the cloud. In this way tenants can \emph{rent} virtual things and get an isolated environment for development and testing. 

Figure~\ref{fig:viriot_system_architecture} shows a high-level view of the VirIoT system architecture. On the left side, there are heterogeneous IoT systems with real things. ThingVisors connect to these and gather information from the real things. This information is processed and exposed as one or more virtual things (VThings). Information related to the vThings are exposed in a common neutral format. For this purpose, the NGSI-LD information representation was chosen that is introduced in Section~\ref{sec:ngsi-ld}. On the right side, virtual silos (vSilos) are shown. Each vSilo consists of a vSilo Controller and an IoT broker. The vSilo Controller is responsible for translating the information from the common neutral format into the target format of the chosen IoT broker and insert it there, ensuring that tenants can access it in the expected format using the API supported by the IoT broker. For example, in the case of oneM2M, the vSilo controller has to create the required REST resource structure consisting of Application Entities, Containers and Content Instances, where the actual information is stored. To configure the VirIoT system, the Master Controller offers a REST API. Through this, all ThingVisors and vSilos are set up. For each vSilo, it has to be decided which vThings are to be added - which could also have a price tag attached and thus would enable a fine-grained charging. For each added vThings, the vSilo has to subscribe to the related information. For communication purposed, an MQTT cluster is used.

The prototype system developed in the Fed4IoT project~\cite{VirIoTPrototype} currently supports oneM2M, FIWARE NGSIv2, NGSI-LD and plain MQTT vSilos. In addition, support for graph databases like Neo4j or ArangoDB, or semantic triple stores like Apache Jena or Eclipse RDF4j could be envisioned.

\begin{figure}[!htbp]
  \centering
  \includegraphics[width=2.5in]{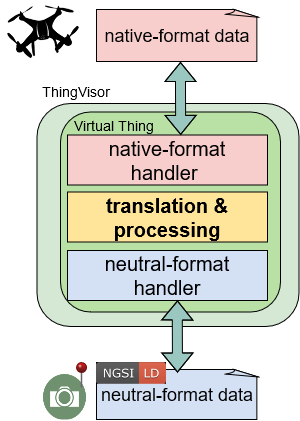}
  \caption{ThingVisor}
  \label{fig:thing_visor}
\end{figure}

Figure~\ref{fig:thing_visor} shows the internal structure of a ThingVisor. It has to support accessing the information from the real thing using the native data format. In the simplest case, it needs to translate this information to NGSI-LD, the chosen neutral format, creating the representation of the corresponding vThing. Whenever there is an update to the vThing, this has to be communicated to the vSilos to which the vThing has been added, using the MQTT communication cluster.
In more advanced cases, the ThingVisor does additional processing, e.g. analyze image frames from a camera, providing the results, e.g. a face count or a recognized face. This information can be mapped to different vThings, i.e. information from one real thing can be mapped to multiple vThings.

If an IoT system provides a huge amount of information, e.g. as in the case of a smart city, implementing the necessary ThingVisors becomes a huge effort and thus it is important to support the developer as much as possible in automating this process. In Section~\ref{sec:ml-based_information_extraction}, we investigate how machine-learning-based information extraction can be used for this purpose.

\section{NGSI-LD as Neutral Exchange Format}
\label{sec:ngsi-ld}

NGSI-LD is an information model~\cite{ngsild2020informationModel} and API~\cite{ngsild2021apispecification} that is being standardized by ETSI, the European Telecommunications Standards Institute, more precisely, by the Industry Specification Group for cross-cutting Context Information Management (ETSI ISG CIM)~\cite{etsiIsgCim2021}.The core idea is to model and manage information about the real world and give applications the context they require for adapting themselves to it.

NGSI-LD is the evolution of what started with the definition of the context interfaces NGSI-9 and NGSI-10~\cite{bauer2010ngsi} by the Open Mobile Alliance (OMA) as part of their Next Generation Service Interfaces (NGSI). The OMA NGSI interfaces were taken up by the Future Internet Public-Private-Partnership in Europe that gave birth to the FIWARE Open Source Commmunity. In FIWARE, the OMA NGSI Context Interfaces were further developed to become NGSIv2 with an HTTP binding and a JSON representation. NGSIv2 became the central interface of the FIWARE open source platform.

In 2017, it was decided to further evolve the NGSI Context Interfaces, which led to the foundation of ETSI ISG CIM. The NGSI-LD information model~\cite{ngsild2020informationModel} got a solid conceptual grounding. It is now based on the property graph model, which allows explicitly modelling entities with properties and relationships and add meta information to them. It is grounded on RDF concepts and supports semantic concept definitions. It follows the linked data idea and uses JSON-LD as its representations format. The NGSI-LD API~\cite{ngsild2021apispecification} enables applications to specify what information they require - based on the NGSI-LD Information Model - including a geographic scoping and temporal interface. In VirIoT, the NGSI-LD API is used for discovery of vThings - which has not been discussed in Section~\ref{sec:iot_virtualization_platform} as it is not in the focus of this paper - and in NGSI-LD vSilos. It is not used for the internal communications, as only a small subset of its functionality would have been needed, and simple MQTT was considered more efficient for this particular functionality. 

\begin{figure*}
  \centering
  \includegraphics[width=6in]{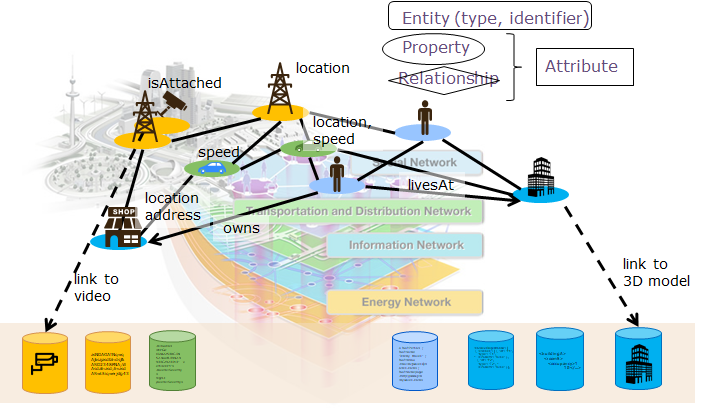}
  \caption{NGSI-LD information model}
  \label{fig:ngsi-ld_information_model}
\end{figure*}

Figure~\ref{fig:ngsi-ld_information_model} shows an example instantiation of the NGSI-LD information model, which by itself can also be seen as a \emph{meta} information model as it defines what kind of elements a fitting domain-specific information model has to define. NGSI-LD models the world as \emph{entities}. Following the linked data principles, each entity has a URI as a unique identifier and one or more entity types. Entity types can be defined as part of a domain-specific information model, sometimes also referred to as a data model. 

In Figure~\ref{fig:ngsi-ld_information_model}, the entity instances shown are two persons, two cars, a building, a shop, two power poles and one camera. Entities can have \emph{properties} and \emph{relationships} - if talking about both, they can be referred to as \emph{attributes}. Which properties and relationships entities of a certain entity type can have, can again be defined in a domain-specific information model. In Figure~\ref{fig:ngsi-ld_information_model}, only a small number of properties are shown, e.g. the shop has a location and an address, and both cars have a speed. Examples for relationships are that one the persons lives at the building and owns the shop. Also, the camera is attached to one of the power poles.

As the example shows, NGSI-LD information represents a property graph with entities as nodes that have properties and relationships as edges. This representation is suitable for information that can easily be attached to a graph structure, but less suitable for bulk information like a 3D model or a stream, e.g. a video stream. In these cases, links into other models stored in suitable storages can be represented as properties, ideally providing all the meta information necessary for accessing the information there. 

As mentioned above, the NGSI-LD information model is really a meta-model that defines only that the world is modelled as entities, properties and relationships, but not which entity types exist and what kind of properties and relationships instances of an entity type can have. This can be done with domain-specific information models. The Smart Data Model initiative, driven by FIWARE, IUDX and TM Forum, is developing the Smart Data Models~\cite{smartdatamodels}, compatible with NGSI-LD, as schemas.The goal is to develop a homogeneous set of models across different IoT domains. 

NGSI-LD was selected as the common neutral format for the VirIoT sytstem, as it is a standard, but also, because on the one hand it is simple enough, on the other hand, it can integrate all the information coming from the source IoT systems in a semantically meaningful way, which may be needed to map it to the information model of the target system in the vSilo.

\section{ML-based Information Extraction}
\label{sec:ml-based_information_extraction}

As we have seen, ThingVisors are the components that translate information from a plethora of information sources to the neutral format NGSI-LD and a conforming information model, a process that can also be called \emph{information extraction}. The source information differs from the target information with respect to the syntactic representation as well as the underlying conceptual model. Such a conceptual model can be explicit or implicit. In the explicit case, a conceptual model is specified in machine-readable form, using some kind of formalism, which can differ with respect to the level of formality. Typical approaches are specifying a schema, vocabulary or ontology. In the implicit case, the conceptual model is directly encoded in the software itself or may be available in textual representation for the programmer, but cannot directly used by a software component.

Programming a ThingVisor constitutes a significant effort. Implementing all ThingVisors required for large IoT systems, like in a Smart City, can be a daunting task. Also, the information models used in the different information silos may be implicit and not readily available. Yet, the more information that can be made available and shared, the better insights can be gained and the more value created by the system. Thus, automating the process or at least making it easier for the programmer, is highly desirable.

The main task in this process is mapping from the source information model to the target information model, the translation of the syntactic representation is then typically straightforward. For the target information model, we assume an integrated backbone model, a common schema or ontology. For NGSI-LD, the Smart Data Models~\cite{smartdatamodels} introduced above form a possible basis for such a backbone model.

\begin{figure}[!htbp]
  \centering
  \includegraphics[width=3.4in]{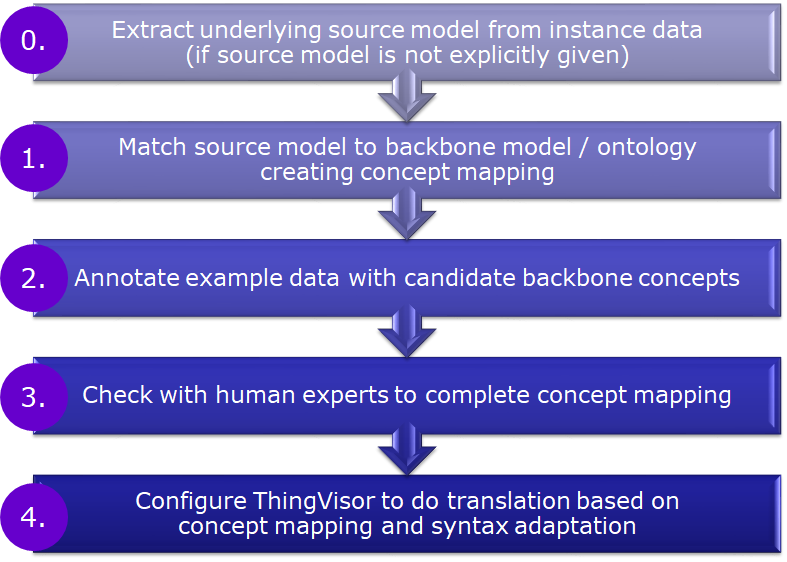}
  \caption{Information extraction steps}
  \label{fig:information_extraction}
\end{figure}

Figure~\ref{fig:information_extraction} shows the different steps for (semi-)automatically creating a concept mapping and information translation that can be used when implementing information extraction in a ThingVisor.

As shown in step 0, if no source model is explicitly given, the attempt can be made to extract a simple, underlying source model from the instance data, taking into account the representation and the structure, e.g. tag names in XML or key names in JSON. In step 1, the source model is matched against the backbone model. This is the most important step of the process as the concepts have to match for the resulting target information to be meaningful. Features derived from structure, naming, and, if available, comments, are used as the basis for this matching. As an example for this step, a machine-learning based approach called knowledge infusion that is used for matching a source ontology to a backbone ontology is shortly presented in Section~\ref{knowledge_infusion}. In step 2, example source data is annotated with candidate backbone concepts in preparation of step 3. There the human expert checks, and if necessary, adapts the mapping, taking into account the annotated source data. In the ideal case, if the automatically created mapping was already perfect, step 3 could be omitted, leading to a fully automated process. However, we expect this to be unrealistic in most real-world cases, and thus the goal is to reduce the required human effort as much as possible. Finally, in step 4, the ThingVisor can be configured to do the translation based on the concept mapping, and the required syntax adaptation, which has to take into account the syntax of the source and the target format.

\subsection{Knowledge Infusion for Ontology Matching}
\label{knowledge_infusion}
Generally, machine learning is a promising approach for many tasks in IoT~\cite{mahdavinejad2018survey}~\cite{solmaz2019pedestrian}
and can also be applied to ontology matching. A typical problem of supervised machine learning approaches is that training a machine learning model requires large amounts of hand-labelled data, which often makes it unrealistic to apply it in many real-world situations. Programmatic labelling, e.g. as used by Snorkel~\cite{ratner2017snorkel}, is a weakly supervised machine learning approach that uses labelling functions to create labelled data, eliminating the need of large amounts of hand-labelled data. Expert knowledge is encoded in the labelling functions, based on which a generative model is built, which is used for labelling data. This in turn is used for training a traditional machine-learning model. Thus, large amounts of labelled data can be made available, leading to better machine-learning models. In~\cite{fuerst2020knowledge} an example is given where the programmatic labelling approach is applied to transport mode detection. For transport mode detection, sensor information from mobile devices such as smartphones is used to determine whether a user is walking, riding a bike, using public transport or driving a car etc. Seven labelling functions were used to create a generative model, which in turn was used for labelling input data and training a random forest machine learning model. This was compared to a random forest model trained based on hand-labelled data. With 80.1\% F1 score, the model trained with with programmatically labelled data achieved almost the same quality as the model trained with hand-labelled data (81.0\%), and significantly better quality than directly using the generative model (74.1\%). This shows that programmatic labelling can lead to similar quality machine learning models than the ones trained with hand-labelled data, but writing a small number of labelling functions can be much cheaper than hand-labelling a large amount of training data.

F\"{u}rst et al.~\cite{fuerst2020knowledge} also describe \emph{knowledge infusion}, a programmatic labelling approach that distinguishes weak and strong knowledge functions. Weak knowledge functions are considered mostly true, whereas strong knowledge functions can be considered as representing canonical truths. Weak and strong knowledge functions are used both as labelling functions in the training face, but are used also, in parallel to the machine-learning model, at execution time, where strong knowledge function can be used to correct wrong outputs and, for example, ensure that safety regulations are always met.

\begin{figure}[!htbp]
  \centering
  \includegraphics[width=3.4in]{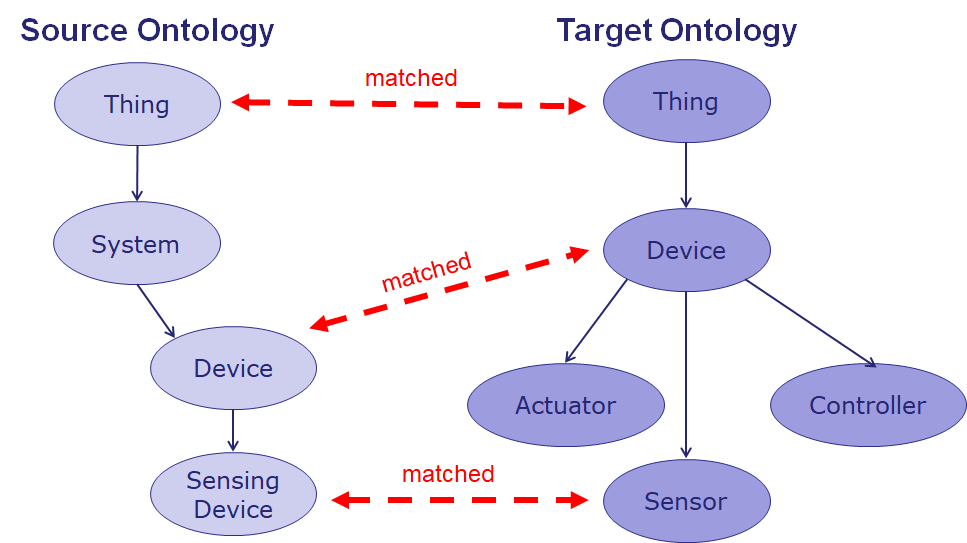}
  \caption{Ontology matching}
  \label{fig:ontology_matching}
\end{figure}

Knowledge infusion can be used in step 1 of Figure~\ref{fig:information_extraction} for matching a source ontology to a target ontology, providing the basis for annotating the data and facilitating the translation to NGSI-LD. Matching in the case of two ontologies means finding the concepts from two ontologies that are equivalent.
Figure~\ref{fig:ontology_matching} shows a simplified example matching three concepts of a source and a target ontology. 

\begin{figure}[!htbp]
  \centering
  \includegraphics[width=3.4in]{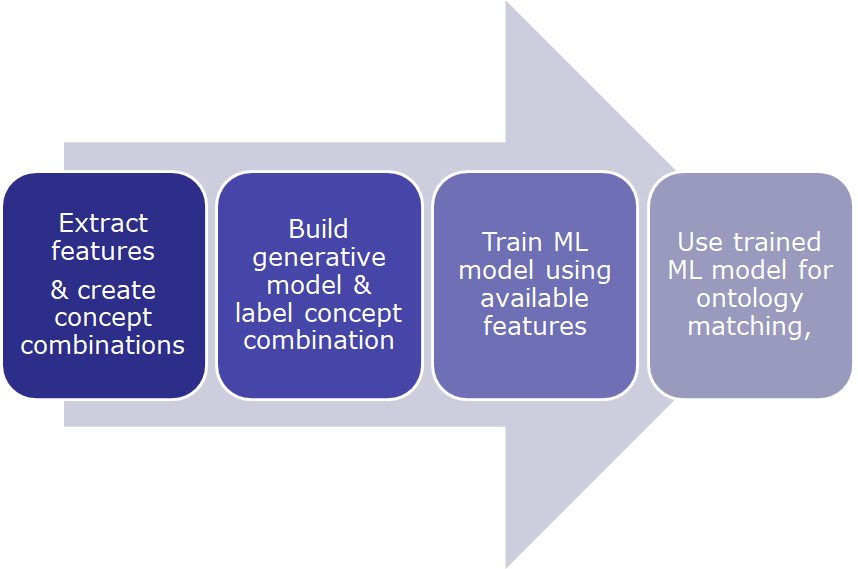}
  \caption{Ontology matching steps}
  \label{fig:ontology_matching_steps}
\end{figure}

Figure~\ref{fig:ontology_matching_steps} shows the different steps required for ontology matching. In the first step, relevant features for ontology concepts are extracted, e.g. concept name, properties, parent and child concepts and a table with all possible combinations of concepts from the source and target ontology are created. In the second step, the generative model is built based on the labelling functions. Examples for labelling functions are matching names, name similarity or synonyms. The generative model is then used to create labels for the concept combinations. With these labelled data, the machine learning model is trained in the third step. Finally, the trained machine learning model is used for ontology matching, i.e. matching concepts from the source and target ontology are identified. Initial experiments with a variety of ontologies show promising results, on a similar or better level than state-of-the-art ontology matching approaches. These will be described in detail in an upcoming paper.

\section{Conclusions and Outlook}
\label{sec:conclusions_and_outlook}

The Internet of Things today is still characterized by a tight coupling between IoT infrastructure deployment and IoT applications. Following the example of device virtualization in the cloud, we have proposed an IoT virtualization architecture that enables the decoupling of IoT infrastructure deployments and IoT applications. This, in turn, enables the sharing of information and more differentiation and specialization in the business roles, because application developers do not have to be infrastructure providers at the same time. 

To build an IoT virtualization platform, an internal common neutral format is needed, because a direct mapping between all source and all target information models is unrealistic. We have introduced the standardized NGSI-LD information model and its characteristics, which make it a suitable choice as a common neutral format.

Furthermore, we have identified that the translation of information from large heterogeneous IoT information systems is labour-intensive, if the code for translation in the ThingVisor needs to be written by hand. We have explored how the developer can be supported using machine learning-based information extraction. As a first step in the process, we have shown initial progress on using machine-learning for ontology matching. This will be further investigated, also taking into account the other steps in the information extraction. The goal is to support the developer with a semi-automatic mapping of concept from the source to the target model, and to further automate the creation of the translation step in ThingVisors. 

\section*{Acknowledgments}
This work is supported in part by the H2020 EU-JP Fed4IoT project and has received funding from the European Union’s Horizon 2020 research and innovation programme under the grant agreement No.~814918 and MIC from Japan. 
The content of this paper does not reflect the official opinion of the European Union. Responsibility for the information and views expressed therein lies entirely with the author.

\bibliographystyle{IEEEtran}
\bibliography{literature}

\end{document}